\documentclass[11pt]{article}  
\usepackage{menuproc}
\usepackage{cite}
\usepackage{epsfig}
\usepackage{amsmath,amssymb}
\begin{document}
%
%
%
\titlematter{Hyperon photo- and electroproduction at CLAS}%
{S.P. Barrow$^a$, representing the CLAS collaboration}%
{$^a$Florida State University\\
     Tallahassee, Florida, 32306\\}
{The large acceptance and high multiplicity capabilities
of the CLAS detector make it possible to study a wide range of
previously unmeasured strange baryon production processes.
Studies of the decay angular distributions of
electroproduced strange baryons have yielded several interesting
new results. The $\Lambda$(1520) electroproduction decay angular
distributions shed light on the spin projections of the $\Lambda$(1520).
Analysis of the decay angular distributions of the weakly decaying
$\Lambda$(1116) have revealed the induced baryon polarization
due to unpolarized incident electron beams.
In addition to these topics, other features of the CLAS
strange baryon program, such as photoproduction and
virtual photon L-T decompositions, are also briefly summarized.}
%
%

\section{Introduction}

The hyperon physics program at CLAS uses polarized
and unpolarized electron beams, 
with incident energies ranging from 2.4 to 6.0 GeV,
to study hyperon photo- and electroproduction.
A complete list of all approved analysis projects currently underway 
is summarized in Table 1.
These experiments measure such aspects of hyperon production as the
production cross sections and the corresponding response functions,
the hyperon decay angular distributions, and the radiative decay strengths
of the light hyperons.

\begin{center}
\begin{tabular}{|l|l|l|}
\hline
Experiment &  Title  & Spokesperson(s) \\
\hline
E89-004 & Hyperon photoproduction & R. Schumacher  \\      
E89-024 & Radiative decays of light hyperons & G. Mutchler \\ 
\hline     
E89-043 & $\Lambda$(1116), $\Lambda$(1520) and $f_{0}$(980) electroproduction & L. Dennis, H. Funsten \\      
E93-030 & Structure functions for kaon electroproduction  & K. Hicks, M. Mestayer \\      
\hline
E95-003 & $K^{\circ}$ electroproduction  & R. Schumacher, K. Dhuga \\      
E99-006 & Polarization observables in $p({\vec e},e'K^{+}){\Lambda}$ & D. Carman, K. Joo,\\ 
 & &  L. Kramer, B. Raue \\    
\hline  
CAA-2000-1 & $K^{*}$ electroproduction & K. Hicks \\      
\hline
\end{tabular}
\end{center}
{\small Table 1. The strange baryon production experiments at CLAS.}

\vspace{2 mm}

\begin{figure}[htb]
\vspace{6.8cm}
\includegraphics{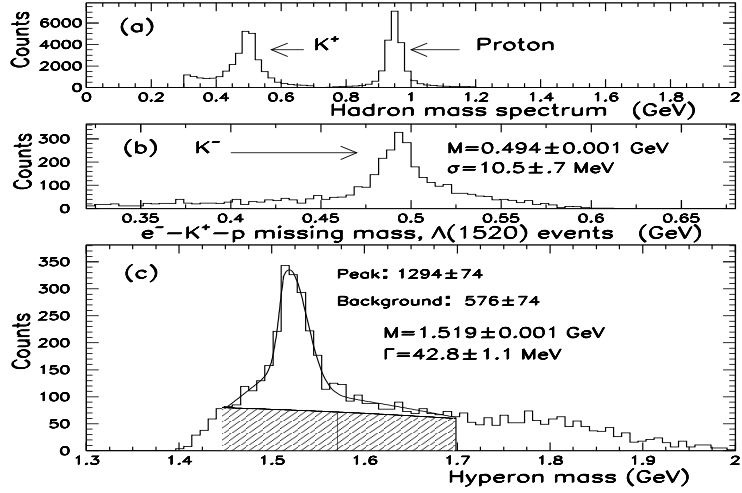}
\caption{ (a) The hadron mass spectrum
for events that contain a proton track and  a $K^+$
candidate.
 (b) The $K^-$ missing mass spectrum for events in which
the e$^-$-$K^+$ missing mass is consistent with the $\Lambda$(1520)
mass. (c) The hyperon mass spectrum for the e$^-$-$K^+$-$K^-$-$p$
final state. A cut on the $K^-$ missing mass from 0.455 to 0.530 GeV was used to generate
this hyperon spectrum.}
\end{figure}

The large acceptance of the CLAS
detector makes it possible to study hyperon production over a wide
kinematic regime. In addition, 
its high multiplicity capabilities enable
the study of sequential processes such
as decay angular distributions of electroproduced hyperons.
Due to time constraints,
the remainder of this talk will focus on the results of measurements of
$\Lambda$(1520) decay angular distributions (E89-043), 
as well as $\Lambda$(1116) decay using unpolarized 
(E89-043) electron beams. 
The kinematic regimes presented here have not
been studied in any previous measurements.


\section{$\Lambda$(1520) decay angular distributions}

The CLAS event reconstruction is based on the missing mass technique
to identify the mass of neutral hyperons and undetected particles.
A study of electroproduction decay angular distributions
requires detecting the scattered electron and at least two
hadrons, and the $\Lambda$(1520) $\rightarrow$
$p-K^{-}$ decay mode of the $\Lambda$(1520) is the one 
best suited for study with CLAS.
Figure 1 shows the relevant missing mass plots for $\Lambda$(1520)
electroproduction data taken as part of the 1998 and 1999 E1 run periods
with beam energies of 4.05, 4.25, and 4.45 GeV.
Reactions that produce other hyperons,
such as the $\Lambda(1405)$, $\Sigma(1480)$, and $\Lambda(1600)$,
account for the majority of the background under the
$\Lambda(1520)$ peak, but
the relative contributions from the
individual processes are currently unknown.
A complete listing of the hyperons whose mass and width have
some overlap with the $\Lambda$(1520) peak 
is presented in Ref. \cite{PDG}.

A  measurement done at Daresbury 
of $\Lambda$(1520) photoproduction \cite{bar80} used incident photons
with energies ranging 
from 2.8 to 4.8 GeV (total
center-of-mass energy $W$ from 2.5 to 3.1 GeV), and 
reports an exponential {\it t}-dependence 
dominated by {\it t}-channel exchange of the $K^{*}$(892) meson,
and not the lighter $K$(494) meson. A thorough understanding of the
reasons $\Lambda$(1520) photoproduction 
proceeds mainly through the exchange
of a heavier vector meson requires theoretical studies
of the 

\begin{figure}[h]
\vspace{5.6cm}
\includegraphics{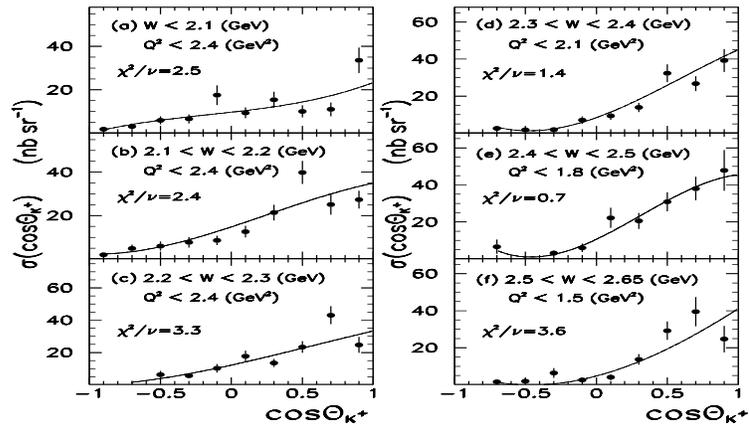}
\caption{The $\Lambda$(1520) $\cos$$\Theta_{K^+}$ differential
cross section distributions for six regions of $W$.
The error bars are statistical uncertainties only.
The solid lines are the results of Legendre polynomial fits to the data.
The lower limit $Q^2$ $=$ 0.9 GeV$^2$ is used for all six distributions.}
\end{figure}

\noindent competition between vector and pseudoscalar meson exchange,
and will not be addressed in this report. However, with CLAS it is possible
to determine if $\Lambda$(1520) electroproduction also proceeds
mainly by {\it t}-channel vector meson exchange. This measurement complements the existing
photoproduction one, and should greatly facilitate a theoretical understanding
of $\Lambda$(1520) production.
The CLAS electroproduction center-of-mass angular distributions shown in Fig.~2 are consistent
with  {\it t}-channel dominance.

\begin{figure}[t]
\vspace{10.6cm}
\includegraphics{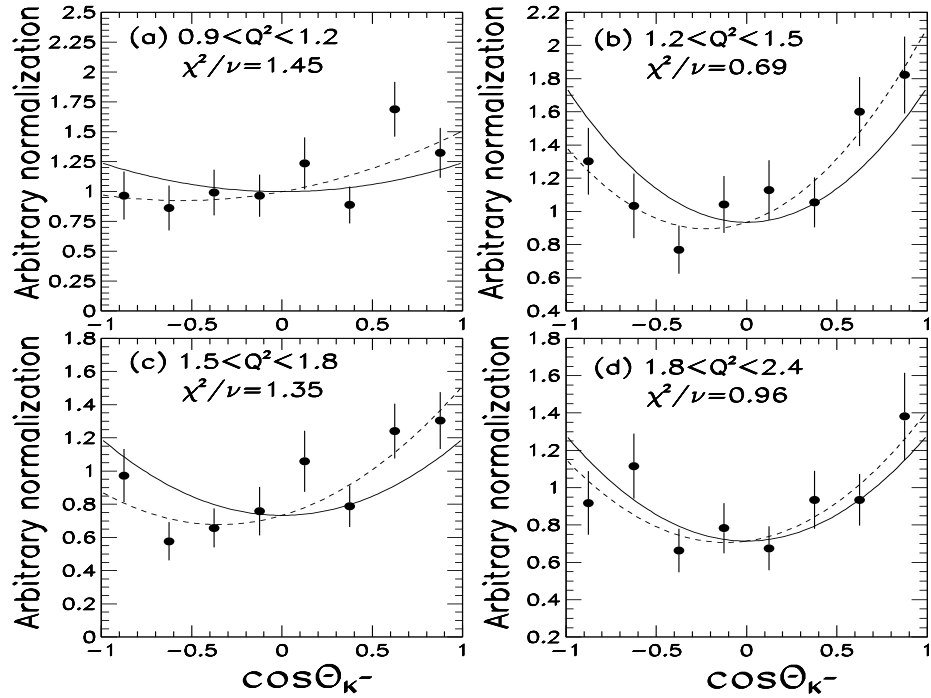}
\caption{\label{pin1}
The $\Lambda$(1520) $\cos$$\Theta_{K^-}$ decay angular
distribution for four regions of $Q^2$. These distributions
are averaged over the region of $W$ from threshold to 2.43 GeV.
The error bars are statistical uncertainties only. The solid line in each
plot is the fitted contribution from the two spin projection terms
of the $\Lambda$(1520), and the dashed line is a fit that also includes
a parameterization of the interference between the $\Lambda$(1520)
and other hyperons.}
\end{figure}

The $\Lambda$(1520) is a J$^{\pi}$ $=$ $\frac{3}{2}^-$ baryon,
and its $p-K^-$ decay is a parity
conserving strong decay mode.
For  an $m_{z}=\pm\frac{3}{2}$ projection the decay 
is characterized by a $\sin$$^{2}\Theta_{K^-}$
distribution, while an $m_{z}=\pm\frac{1}{2}$ projection
has a $\frac{1}{3}$+$\cos$$^{2}\Theta_{K^{-}}$
distribution, where $\Theta_{K^{-}}$ is the polar 
angle of the outgoing $K^{-}$ decay fragment relative to the incident
target proton.
The {\it t}-channel helicity frame $\cos$$\Theta_{K^{-}}$ 
decay angular distributions for four regions
of $Q^2$ are shown in Fig.~3.
The photoproduction angular distribution \cite{bar80} possesses
a greatly enhanced $m_{z}=\pm\frac{3}{2}$ parentage relative to
the electroproduction results presented here. 
All four of the distributions shown in Fig.~3
demonstrate a large
$\frac{1}{3}$+$\cos$$^{2}\Theta_{K^{-}}$ contribution, which indicates
the electroproduced $\Lambda$(1520) hyperons 
are primarily populating the $m_{z}=\pm\frac{1}{2}$ spin projection.

\begin{figure}[t]
\vspace{11.5cm}
\includegraphics{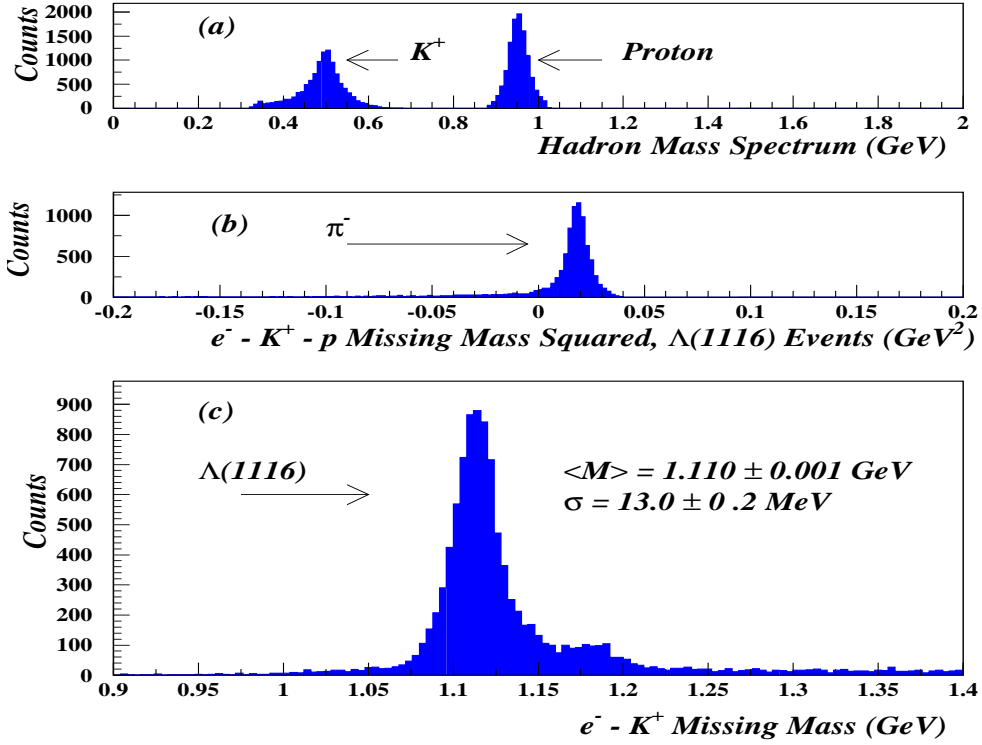}
\caption{\label{pin2}
(a) The hadron mass spectrum
for events that contain a proton track and  a $K^+$
candidate, for $\Lambda$(1116) events.
 (b) The $\pi^-$ missing mass spectrum for events in which
the e$^-$-$K^+$ missing mass is consistent with the $\Lambda$(1116)
mass. (c) The hyperon mass spectrum for the e$^-$-$K^+$-$\pi^-$-$p$
final state.}
\end{figure}

If $\Lambda$(1520) electroproduction proceeds exclusively through {\it t}-channel
exchange of a spinless kaon, the 
$\Lambda$(1520) spin projection is always  $m_{z}=\pm\frac{1}{2}$,
and the ratio  of the $m_{z}=\pm\frac{3}{2}$
to $m_{z}=\pm\frac{1}{2}$ 
populations is zero.
On the other hand, if the reaction  proceeds exclusively through 
the transverse exchange of a J$=$1 $K^*$ vector meson, 
the ratio of the $m_{z}=\pm\frac{3}{2}$
to $m_{z}=\pm\frac{1}{2}$ 
spin projections, if solely determined
by Clebsch-Gordon coefficients, is 3 to 1.
Therefore the  electroproduction
distributions  shown in  Fig.~3, and summarized in Table 2, 
could be evidence for a roughly
equal mixture of $K^*$(892) and $K$(494) contributions, which is a significant
departure from what was reported in the photoproduction measurement \cite{bar80}.
This analysis has recently been published in Phys. Rev. C \cite{spb}.

\begin{center}
\begin{tabular}{|c|c|  }
\hline
{\scriptsize Q$^2$ range (GeV$^2$)} &  {\scriptsize ratio ($m_z=\pm\frac{3}{2})/(m_z=\pm\frac{1}{2}$) }  \\
\hline
0.9-1.2 & .806$\pm$.125  \\      
1.2-1.5 & .534$\pm$.148  \\      
1.5-1.8 & .614$\pm$.108  \\      
1.8-2.4 & .558$\pm$.108  \\      
\hline
\end{tabular}
\end{center}
{\small Table 2. The ratios of the $\Lambda$(1520) electroproduction
spin projection parentages
for the four regions of $Q^2$ presented in Fig.~3. A complete discussion
of these results is presented in Ref. [3] }

\vspace{2 mm}

\begin{figure}[t]
\vspace{10.7cm}
\includegraphics{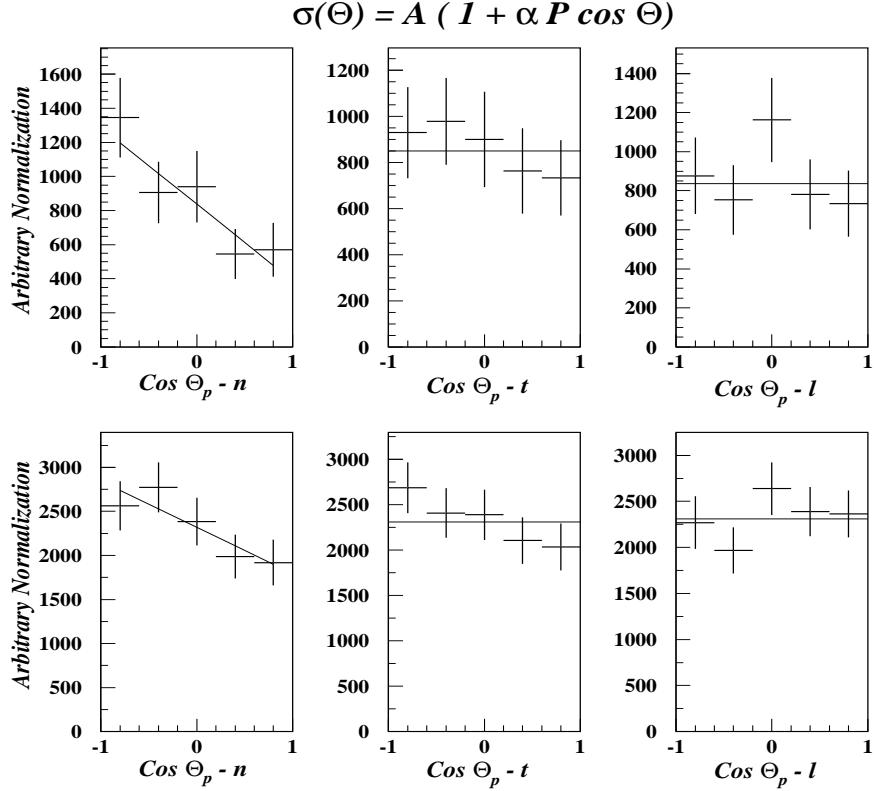}
\caption{\label{pin3}
Two examples of $\Lambda$(1116) decay angular distributions  for
three orthogonal projections. The upper trio is for 1.8 $<$ $W$ $<$ 2.0 GeV
and 0.0 $<$ $Cos\Theta_{p}$ $<$ 0.5, while the lower trio is for the same
range of $W$ but  0.5 $<$ $Cos\Theta_{p}$ $<$ 1.0.
For an unpolarized
incident electron beam, only the ``$n$'' projection, which is
normal to the hyperon production plane, is allowed to
have a slope to its distribution. The ``$t$'' and ``$l$'' projections
are in the center-of-mass production plane, and
should be flat.}
\end{figure}

\section{$\Lambda$(1116) distributions, unpolarized electron beam}

In contrast to the $\Lambda$(1520), the $\Lambda$(1116) is
a J$^{\pi}$ $=$ $\frac{1}{2}^+$ baryon.  
The $\Lambda$(1116) decays weakly, and it is therefore
possible to deduce the polarization of the $\Lambda$(1116) by
studying the asymmetry in its decay angular distribution.
This provides a unique opportunity to study induced baryon polarization 
in hyperon production. Due to parity constraints on the strong interaction,
such polarization is only permitted in the direction
normal to the $\Lambda$(1116) center-of-mass production plane.
The decay angular distribution in the rest frame of the $\Lambda$(1116)
is of the form $\sigma(Cos\Theta_{p}) = A(1+\alpha{\cdot}PCos{\Theta_{p}})$,
where $\Theta_{p}$ is the polar angle of the outgoing proton.
The polarization of the  $\Lambda$(1116) is deduced from the slope
of the $Cos{\Theta_{p}}$ dependence.
The relevant hadron spectra are shown in Fig.~4, and
two examples of the acceptance corrected yields that are used to derive
the $\Lambda$(1116) polarization are shown in Fig.~5.

\begin{figure}[t]
\vspace{10.0cm}
\includegraphics{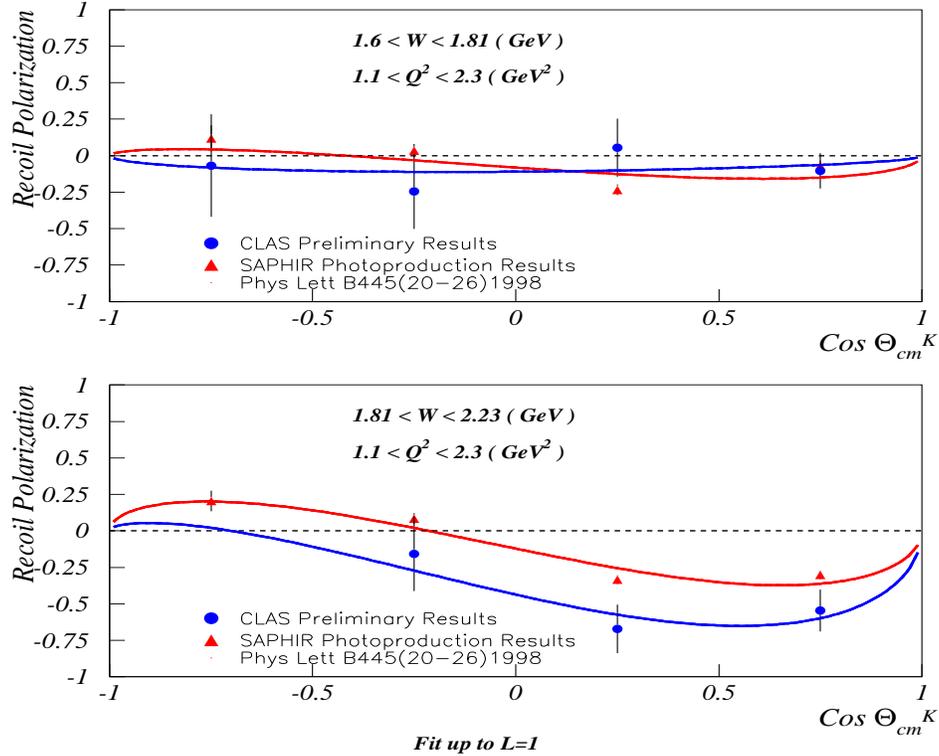}
\caption{\label{pin4}
Preliminary CLAS results for the induced $\Lambda$(1116) polarization
from an unpolarized incident electron beam for two regions of $W$, 
and for 1.1 $<$ $Q^2$ $<$ 2.3 GeV$^2$. Also shown in these plots
are the results of a photoproduction measurement [5].}
\end{figure}


The induced polarization as a function of the center-of-mass
quantity $\cos$$\Theta_{K^+}$
for two regions of $W$ are shown in Fig.~6 \cite{si}.
Also shown are the results of a photoproduction measurement
done at SAPHIR \cite{saphir} 
of $\Lambda$(1116) induced polarization over the same region of $W$
studied with CLAS. The photo- and electroproduction measurements both
indicate that for low $W$, close to threshold, the induced polarization
of the $\Lambda$(1116) is fairly small, while at higher $W$
the $\Lambda$(1116) polarization is larger and negative, especially
for $Cos${$\Theta_{K^+}$} $>$ 0.0. The photo- and electroproduction
data also suggest there might be no 
significant  $Q^2$ dependence to the induced $\Lambda$(1116) polarization.
One obvious implication of this result is that the $\Lambda$(1116) polarization
is not very sensitive to the L-T decomposition of the (real or virtual) photon.
Once this analysis is complete, it will be 
interesting to compare the $W$ dependence of the $\Lambda$(1116) 
induced polarization
presented here with the polarization  obtained with a 
nonlepton beam such
as a pion or kaon beam.

\section{Conclusions}

The detailed studies
of the decay angular distributions of electroproduced
hyperons presented here represent a significant addition to existing
measurements of hyperon production. 
These measurements also provide excellent illustrations
of the capabilities of the CLAS detector.
Nonetheless, these results represent a small fraction of
the total studies of hyperon production currently
underway on data taken at CLAS. Combined with
the other approved analysis activities summarized in Table 1, 
a much clearer understanding
of the strange baryon production  processes is emerging.

\acknowledgments{
The CLAS collaboration is supported by the U.S. Department of Energy and the
National Science Foundation, the French Commissariat \`{a} l'Energie
Atomique, the Italian Istituto Nazionale di Fisica Nucleare, and
the Korea Science and Engineering Foundation. }



\begin{references}[9]   
\bibitem{PDG} D. Groom $et$ $al$., The European Physical Journal C {\bf 15}, 1 (2000).
\bibitem{bar80} D. Barber $et$ $al.$, Z. Physik C
{\bf 7}, 17 (1980).
\bibitem{spb} S.P. Barrow $et$ $al.,$ Phys. Rev. C {\bf 64}, 044601 (2001), hep-ex/0105029.
\bibitem{si} Data analysis results from S. McAleer (2001).
\bibitem{saphir} M. Q. Tran  $et$ $al.$, Phys.
Lett. B {\bf 445}, 20 (1998).

\end{references}
\end{document}